\documentclass[amsmath,superscriptaddress,aps,prb,onecolumn]{revtex4}
\usepackage[colorlinks,linkcolor=blue,anchorcolor=blue,citecolor=blue,urlcolor=blue]{hyperref}
\usepackage{amsmath}
\usepackage{amssymb}
\usepackage{graphicx}
\usepackage{color}
\usepackage{bm}

\begin{document}
\title{The Analysis of Bulk Boundary Correspondence under the Singularity of the Generalized Brillouin Zone in Non-Hermitian System}

\author{Gang-Feng Guo}
\affiliation{Lanzhou Center for Theoretical Physics, Key Laboratory of Theoretical Physics of Gansu Province, Lanzhou University, Lanzhou, Gansu 730000, China}
\affiliation{Key Laboratory for Magnetism and Magnetic Materials of the Ministry of Education, Lanzhou University, Lanzhou 730000, People's Republic of China}

\author{ Xi-Xi Bao}
\affiliation{Lanzhou Center for Theoretical Physics, Key Laboratory of Theoretical Physics of Gansu Province, Lanzhou University, Lanzhou, Gansu 730000, China}
\affiliation{Key Laboratory for Magnetism and Magnetic Materials of the Ministry of Education, Lanzhou University, Lanzhou 730000, People's Republic of China}


\author{Lei Tan}
\email{tanlei@lzu.edu.cn}
\affiliation{Lanzhou Center for Theoretical Physics, Key Laboratory of Theoretical Physics of Gansu Province, Lanzhou University, Lanzhou, Gansu 730000, China}
\affiliation{Key Laboratory for Magnetism and Magnetic Materials of the Ministry of Education, Lanzhou University, Lanzhou 730000, People's Republic of China}

\begin{abstract}
The generalized Brillouin zone (GBZ), which is the core concept of the non-Bloch band theory to rebuild the bulk boundary correspondence in the non-Hermitian topology, appears as a closed loop generally. In this work, we find that even if the GBZ itself collapses into a point, the recovery of the open boundary energy spectrum by the continuum bands remains unchanged. Contrastively, if the bizarreness of the GBZ occurs, the winding number will become illness. Namely, we find that the bulk boundary correspondence can still be established whereas the GBZ has singularities from the perspective of the energy, but not from the topological invariants. Meanwhile, regardless of the fact that the GBZ comes out with the closed loop, the bulk boundary correspondence can not be well characterized yet because of the ill-definition of the topological number. Here, the results obtained may be useful for improving the existing non-Bloch band theory.

\end{abstract}

\maketitle
\section{INTRODUCTION}
Recently, the topological insulators in the Hermitian system has attracted considerable research interest [\onlinecite{Hasan1, Qi2, Bansil3, Armitage4, Chiu5, 6, 7, 8, 9, 10}]. The bulk boundary correspondence is among the central concepts and has been established perfectly using the Bloch band theory [\onlinecite{11, 12, 13, 14, 15}], in which the boundary refers to the energy spectrum originating from the open boundary condition. The bulk contains two physical meanings, one of which is the energy spectrum calculated by the periodic boundary condition, while another one senses the topological invariant of the system calculated in the momentum space as well. Universally, the physical elucidation of the bulk boundary correspondence is that the energy spectrum under the periodic boundary and the one under the open boundary coincides with each other ideally at the thermodynamic limit, except for the zero mode being predicted by the topological invariant.

However, the non-Hermitian topological systems, which introduces some dissipative ingredients, have extended the frontier of the law of physics and raised many novel phenomena [\onlinecite{16, 161, 162, 163, 164, 165, 166, 167, 17, 18, 19, 20, 21, 22, 23, 24, 27, 28, 29, 30, 32, 33, 34, 35, 37, 38, 39, 40, 41, 42,43, 44, 45, 46, 47, 48, 49, 50}]. Among the key aspects is the nullities of the bulk boundary correspondence in the non-Hermitian systems [\onlinecite{16, 161, 162, 163, 164, 165}]. Concretely, the spectra under different boundary conditions have an overt distinction [\onlinecite{16, 161, 162, 163, 164, 165, 166, 167}]. To understand the phenomenon deeply, the Ref. [\onlinecite{163}] extended the Bloch wave vector from the real number to the complex number creatively and thus introduced a new concept of the Generalized Brillouin zone (GBZ), which is the heart of the non-Bloch band theory [\onlinecite{163, 164, 165, 166, 167}]. The information read off from the GBZ includes that it is a closed curve generally and the cusps will appear in some cases [\onlinecite{163, 164, 165, 166, 167}]. Depending on the GBZ, the open boundary energy spectrum can be reproduced by the continuum bands, and further, the zero energy modes also can be forecasted by the non-Bloch topological invariant. Namely, the bulk boundary correspondence in the non-Hermitian topology is reinstituted from the aspects of the energy and topological number relied on the GBZ. But, we must emphasize that, based on the current framework of the non-Bloch band theory, the recovery of the bulk boundary correspondence requires that GBZ is a closed curve in general [\onlinecite{163, 164, 165, 166, 167}]. By contrast, what will happen if the GBZ collapses into a point? Precisely speaking, what effect does the singularity of the GBZ have on the bulk boundary correspondence of the system? In addition, while the GBZ is a closed curve as we expected, will the bulk boundary correspondence hold for granted?

In this work, we explore these important questions by considering a non-Hermitian Su-Schrieffer-Heeger model since it possesses the structural simplicity and the abundant physical insight concurrently to clarify the impact of GBZ's bizarre nature on the bulk boundary correspondence. It can be shown that whether the GBZ is a closed curve or not, the energy spectrum in an open chain always can be regained by the continuum bands faithfully, which illustrates that the bulk boundary correspondence is still valid from the energy side. On the other hand, the topological invariant will not be well-defined provided the collapse occurs to the GBZ, which demonstrates that the bulk boundary correspondence is illness from the topological invariant aspect. It also can be found that whereas the GBZ emerges as a closed curve, the bulk boundary correspondence will not be established definitely.

The paper is organized as follows. Sec. \ref{II} provides the theoretical model and introduces the concepts of the GBZ, the relevant topological invariant and continuum bands. Sec. \ref{III} is devoted to analyse the effect of the GBZ being a point on the bulk boundary correspondence. Sec. \ref{IV} researches the situation where the bulk boundary correspondence is still illness even when the GBZ exists as a closed loop. Finally, the conclusions are showed in Sec. \ref{V}.

\section{MODEL AND THEORY}\label{II}
We consider a one-dimensional non-Hermitian SSH system, which can be described by
\begin{flalign}
H=&\sum_{n}\Big[(t_{1}+\frac{\gamma_{1}}{2}){C^\dag}_{A,n}{C}_{B,n}+(t_{1}-\frac{\gamma_{1}}{2}){C^\dag}_{B,n}{C}_{A,n}+\nonumber\\
&(t_{2}+\frac{\gamma_{2}}{2}){C^\dag}_{B,n}{C}_{A,n+1}+(t_{2}-\frac{\gamma_{2}}{2}){C^\dag}_{A,n+1}{C}_{B,n}+\nonumber\\
&(t_{3}+\frac{\gamma_{3}}{2}){C^\dag}_{A,n}{C}_{B,n+1}+(t_{3}-\frac{\gamma_{3}}{2}){C^\dag}_{B,n+1}{C}_{A,n}\Big] ,
\end{flalign}
where ${C^\dag}_{An},_{Bn}$ (${C}_{An},_{Bn}$) is the creation (annihilation) operator on the sublattices $A$, $B$ in the nth unit cell. $\gamma_{1}$, $\gamma_{2}$ and $\gamma_{3}$ represent the non-Hermiticity parameters. $t_{1}$, $t_{2}$ and $t_{3}$ characterize the intracell and intercell hoppings.

Based on the Schr\"{o}dinger equation $H|\psi\rangle=E|\psi\rangle$, where $|\psi\rangle=(\psi_{A,1},\psi_{B,1},...,\psi_{A,n},\psi_{B,n},...)^{T}$, the eigen equation in the bulk will be written as
\begin{equation}
(t_{2}-\frac{\gamma_{2}}{2})\psi_{B,n}+(t_{1}+\frac{\gamma_{1}}{2})\psi_{B,n+1}+(t_{3}+\frac{\gamma_{3}}{2})\psi_{B,n+2}=E\psi_{A,n+1},
\end{equation}

\begin{equation}
(t_{3}-\frac{\gamma_{3}}{2})\psi_{A,n}+(t_{1}-\frac{\gamma_{1}}{2})\psi_{A,n+1}+(t_{2}+\frac{\gamma_{2}}{2})\psi_{A,n+2}=E\psi_{B,n+1}.
\end{equation}
The elements of the wavefunction have the form $\psi_{A,n}=\beta^{n} \phi_{A}$ and $\psi_{B,n}=\beta^{n} \phi_{B}$ [\onlinecite{163}, \onlinecite{165}]. Then, one can get
\begin{equation}
(t_{2}-\frac{\gamma_{2}}{2})\phi_{B}+(t_{1}+\frac{\gamma_{1}}{2})\beta\phi_{B}+(t_{3}+\frac{\gamma_{3}}{2})\beta^{2}\phi_{B}=E\beta\phi_{A},\label{4}
\end{equation}

\begin{equation}
(t_{3}-\frac{\gamma_{3}}{2})\phi_{A}+(t_{1}-\frac{\gamma_{1}}{2})\beta\phi_{A}+(t_{2}+\frac{\gamma_{2}}{2})\beta^{2}\phi_{A}=E\beta\phi_{B}.\label{5}
\end{equation}
Hence, the quadratic equation about $\beta$ can be obtained and the GBZ can be generally determined by $\beta_{2}$ and $\beta_{3}$ satisfying $\left|\beta_{1}\right|\leq\left|\beta_{2}\right|=\left|\beta_{3}\right|\leq\left|\beta_{4}\right|$, which are the solutions of this eigenvalue equation for a given $E$.

In addition, from the real space Hamiltonian, the generalized Bloch Hamiltonian also can be acquired as
\begin{equation}       
H(\beta)=\left(                 
  \begin{array}{ccc}   
    0 & R_+(\beta) \\  
    R_-(\beta) & 0 \\  
  \end{array}
\right),                 
\end{equation}
where $H(\beta)$ is the counterpart of the Bloch Hamiltonian $H(k)$ through replacing $e^{ik}\rightarrow\beta$, where $k$ $\in$ $\mathbb{C}$. Thus, the continuum bands and the topological invariant can be defined using the ingredients of $H(\beta)$ as [\onlinecite{163}, \onlinecite{165}]
\begin{equation}
E^{2}_{GBZ}=R_{+}(\beta_{GBZ})R_{-}(\beta_{GBZ}),
\end{equation}
\begin{equation}
W=-\frac{\big(\arg R_{+}(\beta)-\arg R_{-}(\beta)\big)_{GBZ}}{4\pi}.\label{8}
\end{equation}

From the definition Eq. \eqref{8}, the geometrical meaning of the topological invariant can be explained as the change of the phase of $R_{\pm}(\beta)$ with $\beta$ turning around the GBZ in the counterclockwise direction. In other words, the topological order can be simply ensured by counting how many times the $R_{\pm}(\beta)$ contains the origin.


Next, based on the topological invariant and continuum bands associating with the GBZ, and the energy spectrum under different boundary conditions, the bulk boundary correspondence will be demonstrated from the signatures of the GBZ.

\section{the effect of The Generalized Brillouin Zone being a point on the Bulk Boundary Correspondence}\label{III}
In this section, we will talk about the impact on the bulk boundary correspondence when GBZ collapses into a point, and compare the results to the normal case.
\begin{footnotesize}
\subsection{\normalsize The GBZ Formed By $\beta_{1}$ And $\beta_{2}$ \normalsize}
\end{footnotesize}

For simplicity, $t_{3}=\gamma_{3}=0$ can be explored firstly. Here, the analytical expressions of the GBZ and the corresponding continuum bands $E_{GBZ}$ can be easily achieved.

\begin{figure}[!htbp]
\includegraphics[width=4.2cm,height=3.8cm]{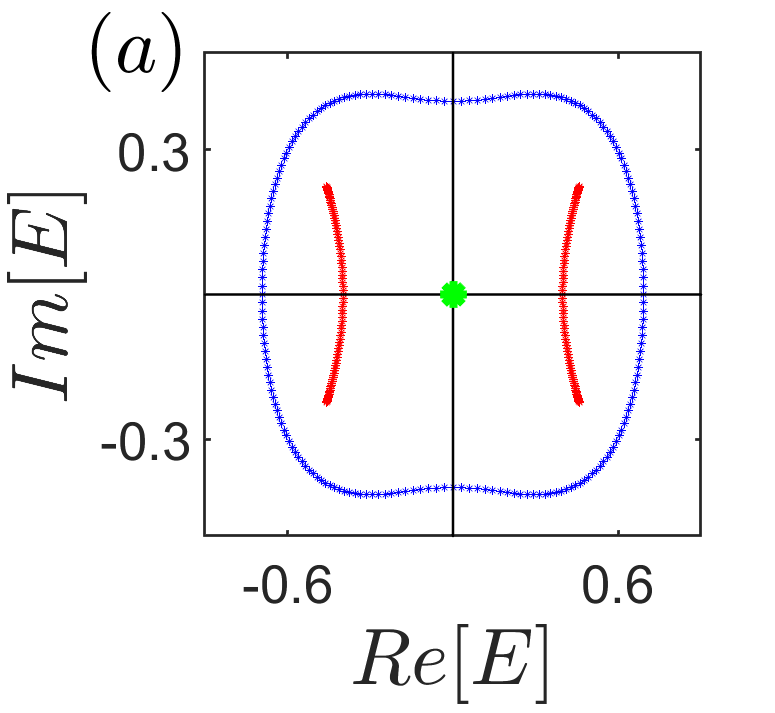}
\includegraphics[width=4.2cm,height=3.8cm]{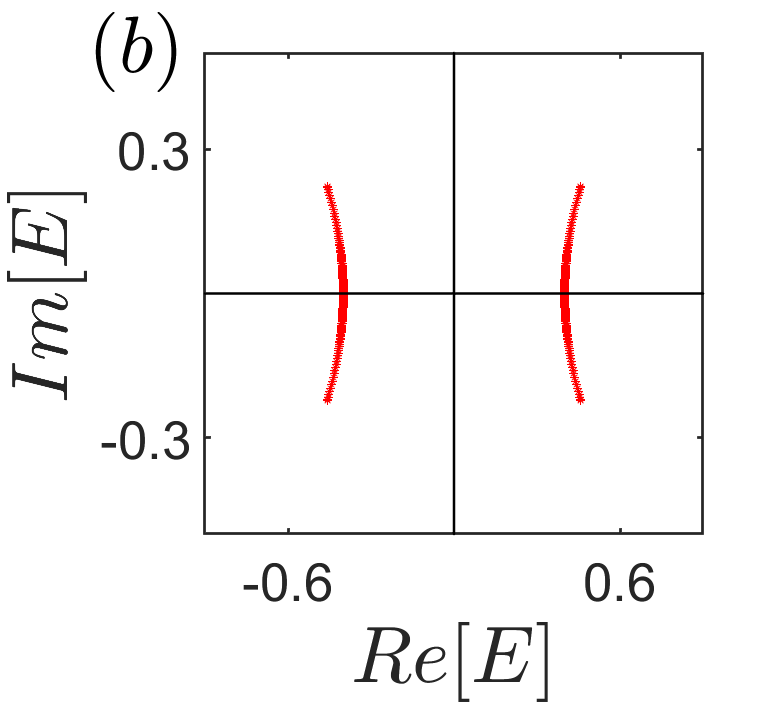}

\includegraphics[width=4.2cm,height=3.8cm]{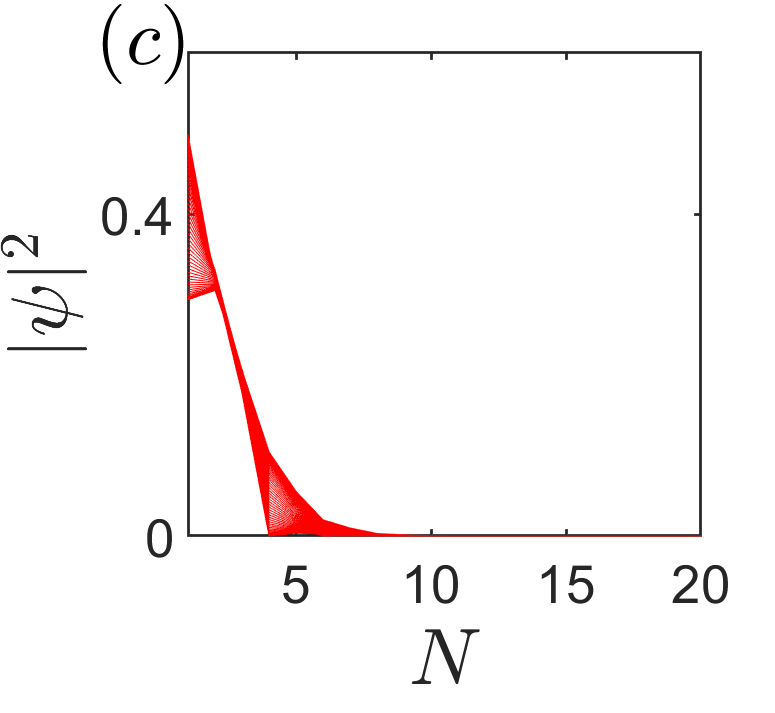}
\includegraphics[width=4.2cm,height=3.8cm]{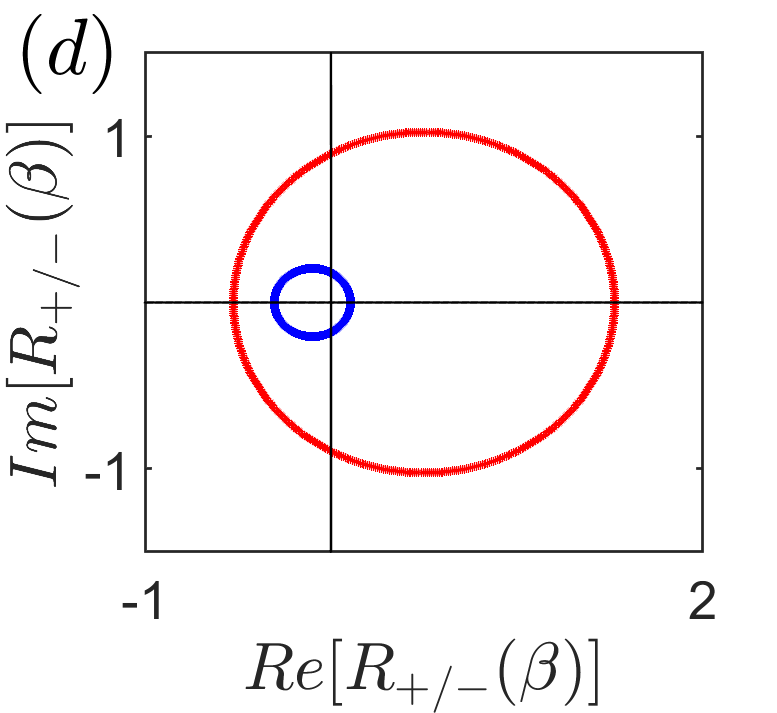}
\caption{(Color online) $N=100$, $\gamma_{1}=0.6$, $\gamma_{2}=0.4$, $\gamma_{3}=0$, $t_{1}$=0.2, $t_{2}=0.5$, $t_{3}=0$. (a) The energy spectrum in open boundary condition (the red line) and in periodic boundary condition (the blue line), respectively. The green dot denotes the two-degenerate zero modes of the open chian. (b) The continuum bands, which is same as the open boundary spectrum. (c) The skin modes of all the bulk eigenstates. For clarity, we truncate $N$ = $20$. (d) The loops of $R_{+}(\beta)$ (the red line) and $R_{-}(\beta)$ (the blue line). Apparently, the origin is included by both $R_{+}(\beta)$ and $R_{-}(\beta)$, which infers $W=1$.}
\label{fig1}
\end{figure}

According to Eqs. \eqref{4} and \eqref{5}, the characteristic equation of $\beta$ are
\begin{equation}
[(t_{1}-\frac{\gamma_{1}}{2})+(t_{2}+\frac{\gamma_{2}}{2})\beta][(t_{2}-\frac{\gamma_{2}}{2})+(t_{1}+\frac{\gamma_{1}}{2})\beta]=E^{2}\beta.
\end{equation}
The two solutions have the form $\beta_{1,2}=\frac{-b\pm\sqrt{b^{2}-4ac}}{2a}$ with $a=(t_{1}+\frac{\gamma_{1}}{2})(t_{2}+\frac{\gamma_{2}}{2})$, $b=(t_{1}-\frac{\gamma_{1}}{2})(t_{1}+\frac{\gamma_{1}}{2})+(t_{2}-\frac{\gamma_{2}}{2})(t_{2}+\frac{\gamma_{2}}{2})-E^{2}$ and $c=(t_{1}-\frac{\gamma_{1}}{2})(t_{2}-\frac{\gamma_{2}}{2})$. The discussion above applies, and therefore the trajectories of $\beta_{1}$ and $\beta_{2}$ satisfying $|\beta_{1}|=|\beta_{2}|$ constitute the GBZ, which implies $\sqrt{b^{2}-4ac}=-i \eta b$, $\eta\in \mathbb R$ [\onlinecite{51}, \onlinecite{52}]. Hence, the GBZ and continuum bands can be described by

\begin{equation}
|\beta_{GBZ}|=\sqrt{\frac{|(t_{1}-\frac{\gamma_{1}}{2})(t_{2}-\frac{\gamma_{2}}{2})|}{|(t_{1}+\frac{\gamma_{1}}{2})(t_{2}+\frac{\gamma_{2}}{2})|}},\label{10}
\end{equation}
and
\begin{equation}
E^{2}_{GBZ}=(t_{1}+\frac{\gamma_{1}}{2})(t_{1}-\frac{\gamma_{1}}{2})+(t_{2}+\frac{\gamma_{2}}{2})(t_{2}-\frac{\gamma_{2}}{2})\mp\sqrt{\frac{4ac}{1+\eta^{2}}}.
\end{equation}
Eq. \eqref{10} clearly shows that the GBZ is a circle with the radius of $|\beta_{GBZ}|$, which is undoubtedly a closed loop in most cases. As shown in Fig. \ref{fig1}(a), the energy spectrum in open boundary condition composed of the red line and the green dot is deviated from the one in periodic boundary condition (the blue line), but is consistent with the continuum bands in Fig. \ref{fig1}(b) calculated through $E^{2}_{GBZ}$ except for the green dot standing for the two-degenerate zero modes. This deviation originates from the non-Hermitian skin modes [\onlinecite{163}, \onlinecite{53}], illustrated schematically in Fig. \ref{fig1}(c).

On the other hand, the expressions $R_{+}(\beta)$ and $R_{-}(\beta)$ can be acquired as
\begin{equation}
R_{+}(\beta) = t_{1}+\frac{\gamma_{1}}{2}+(t_{2}-\frac{\gamma_{2}}{2})\beta^{-1},
\end{equation}
\begin{equation}
R_{-}(\beta) = t_{1}-\frac{\gamma_{1}}{2}+(t_{2}+\frac{\gamma_{2}}{2})\beta.
\end{equation}
As shown in Fig. \ref{fig1}(d), both the images of $R_{+}(\beta)$ and $R_{-}(\beta)$ encircle the origin once when $\beta$ goes along the GBZ in the counterclockwise way, which means the system is in the nontrivial phase with $W$=$1$ and has the zero modes corresponding the green dot in Fig. \ref{fig1}(a).

\begin{figure}[!htbp]
\includegraphics[width=4.2cm,height=3.8cm]{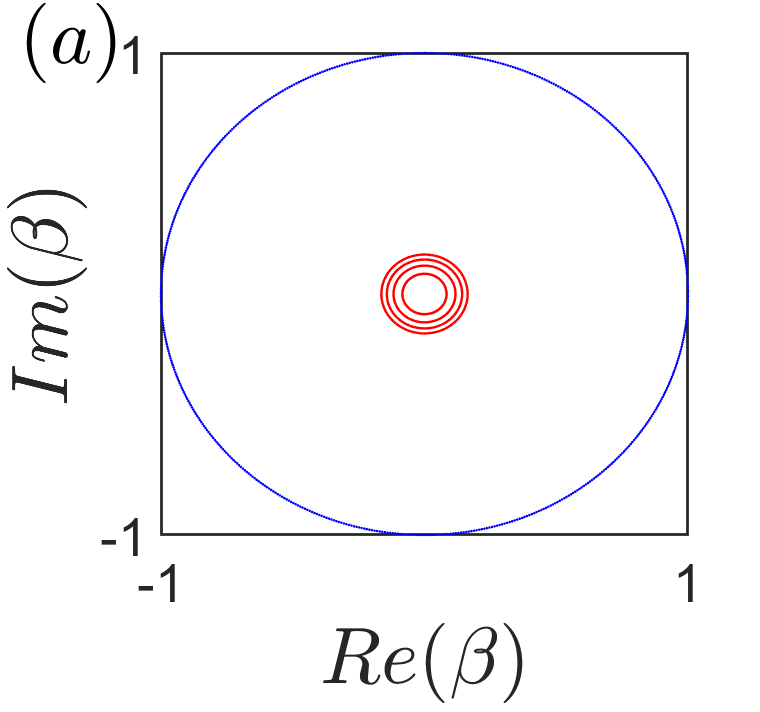}
\includegraphics[width=4.2cm,height=3.8cm]{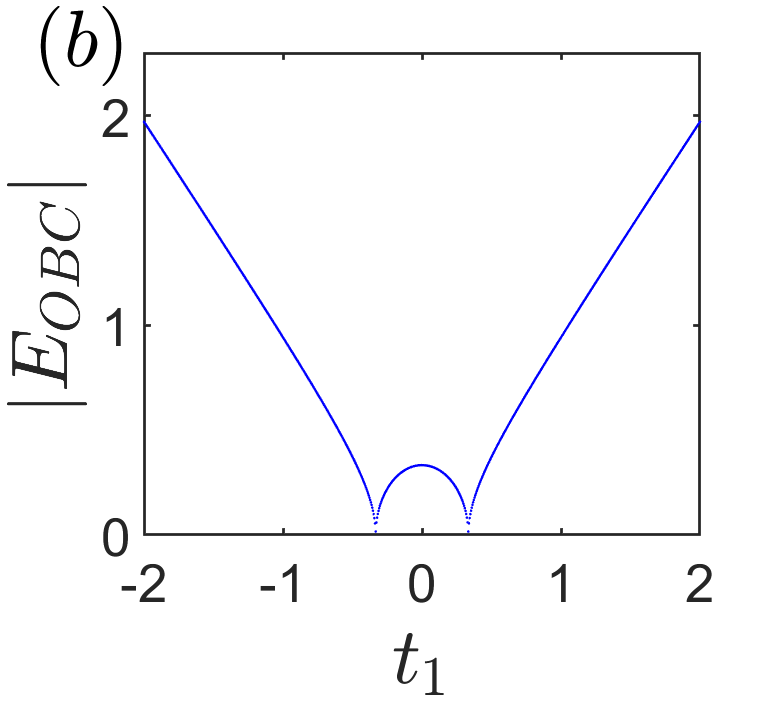}

\includegraphics[width=4.2cm,height=3.8cm]{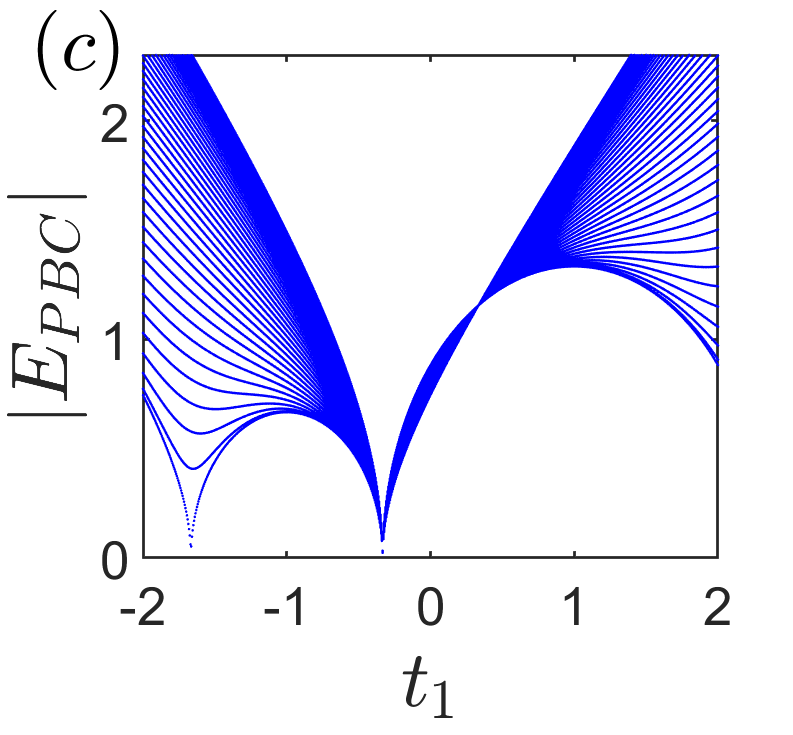}
\includegraphics[width=4.2cm,height=3.8cm]{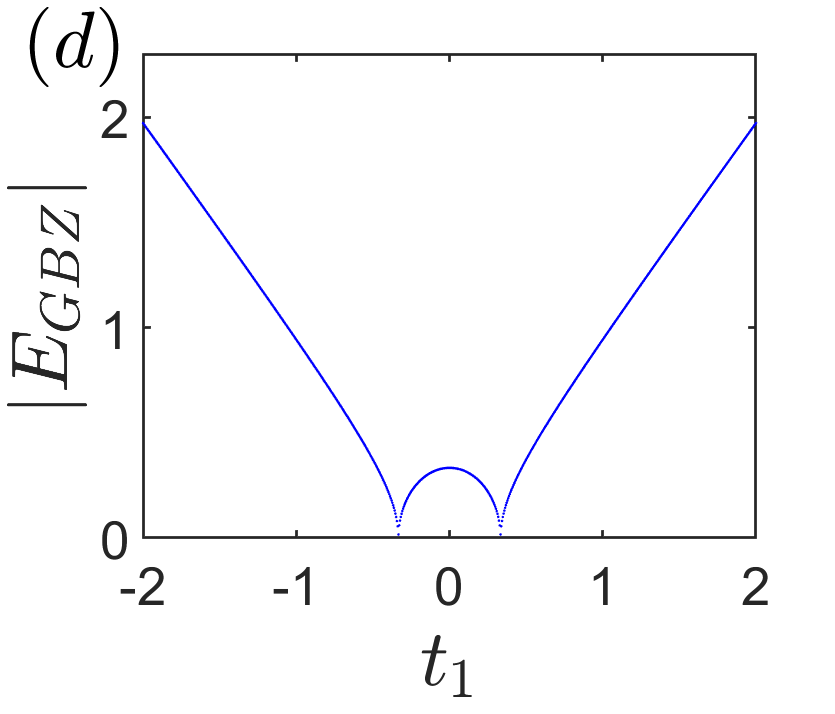}

\caption{(Color online) (a) The green line being the Brillouin zone, while the red lines being the GBZ. From outside to inside, $t_{1}=0.34, 0.33, 0.32, 0.31$, respectively. Obviously, when $t_{1}$ closes to $\frac{\gamma_{1}}{2}$= $0.3$, the GBZ develops a tendency to collapse into a point. The other parameters are same as the Fig. 1. For (b), (c) and (d), $N=100$, $\gamma_{1}=\frac{2}{3}$, $\gamma_{3}=0$, $t_{3}=0$ and $t_{2}=\frac{\gamma_{2}}{2}=1$. (b) The open boundary energy spectrum, which matches well with the continuum bands of (d), but deflects the energy spectrum under periodic boundary condition in (c).}
\label{fig2}
\end{figure}

The analysis process above shows that the recovery of the bulk boundary correspondence relies on both $R_{+}(\beta)$ and $R_{-}(\beta)$ as closed curves, which requires that GBZ itself must be a closed curve as well. However, we notice that if either $t_{1}$ closing to $\frac{\gamma_{1}}{2}$ or $t_{2}$ closing to $\frac{\gamma_{2}}{2}$, the GBZ will deforms into a point gradually, as indicated in Fig. \ref{fig2}(a), which implies that the GBZ has the singular features.

To investigate the singularities of the GBZ more visually, the parameters can be $t_{2}$ = $\frac{\gamma_{2}}{2}$ =$1$ in Eq. \eqref{10}, by which GBZ becomes a point. The violation in Figs. \ref{fig2}(b) and \ref{fig2}(c) presents that the open boundary energies can not be reproduced by the periodic boundary energies yet. Amusingly, even if the GBZ collapses into a point, the open boundary energy spectrum still coincides with the continuum bands $E_{GBZ}$, as shown in Figs. \ref{fig2}(b) and \ref{fig2}(d). However, the winding number is ill-defined because both $R_{+}(\beta)$ and $R_{-}(\beta)$ not emerge as the closed loop, which is attributed to the singularities of the GBZ. Above all, under the condition that the GBZ has strangeness, the bulk boundary correspondence is still valid from the perspective of energy spectrum, but not from the side of topological invariants, i.e., the correctness of the bulk boundary correspondence has been destroyed partially with the GBZ being a point. 

\begin{footnotesize}
\subsection{\normalsize The GBZ formed by $\beta_{2}$ and $\beta_{3}$ \normalsize}
\end{footnotesize}
Next, the case $t_{2}=\frac{\gamma_{2}}{2}$ can be analyzed. The characteristic equation of $\beta$ has the form
\begin{equation}
[(t_{1}+\frac{{\gamma_{1}}}{2})\beta+(t_{3}+\frac{{\gamma_{3}}}{2})\beta^{2}][(t_{1}-\frac{{\gamma_{1}}}{2})\beta+2t_{2}\beta^{2}+(t_{3}-\frac{{\gamma_{3}}}{2})]=E^{2}\beta^{2}.\label{14}
\end{equation}
The parameters given ensure that this is a quartic equation. The solutions can be ordered as $\left|\beta_{1}\right|\leq\left|\beta_{2}\right|\leq\left|\beta_{3}\right|\leq\left|\beta_{4}\right|$
for the eigenvalue E and only $\left|\beta_{2}\right|=\left|\beta_{3}\right|$ can recover the bulk boundary correspondence.

\begin{figure}[!htbp]
\includegraphics[width=4.2cm,height=3.8cm]{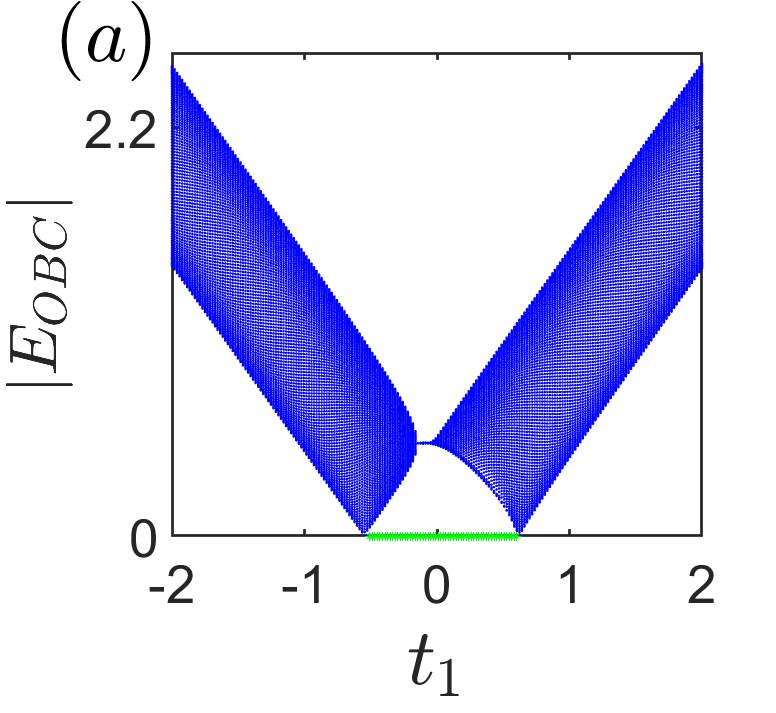}
\includegraphics[width=4.2cm,height=3.8cm]{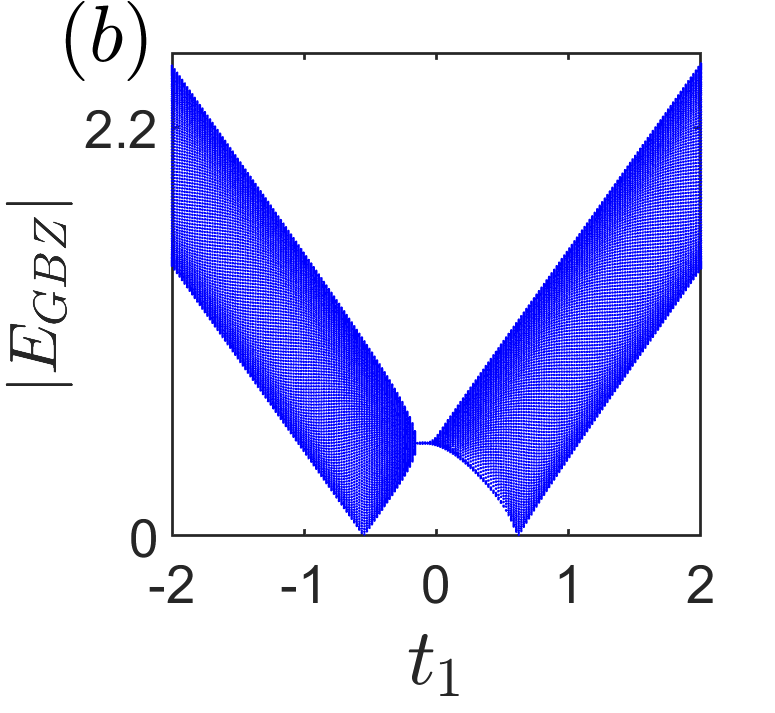}

\includegraphics[width=4.2cm,height=3.8cm]{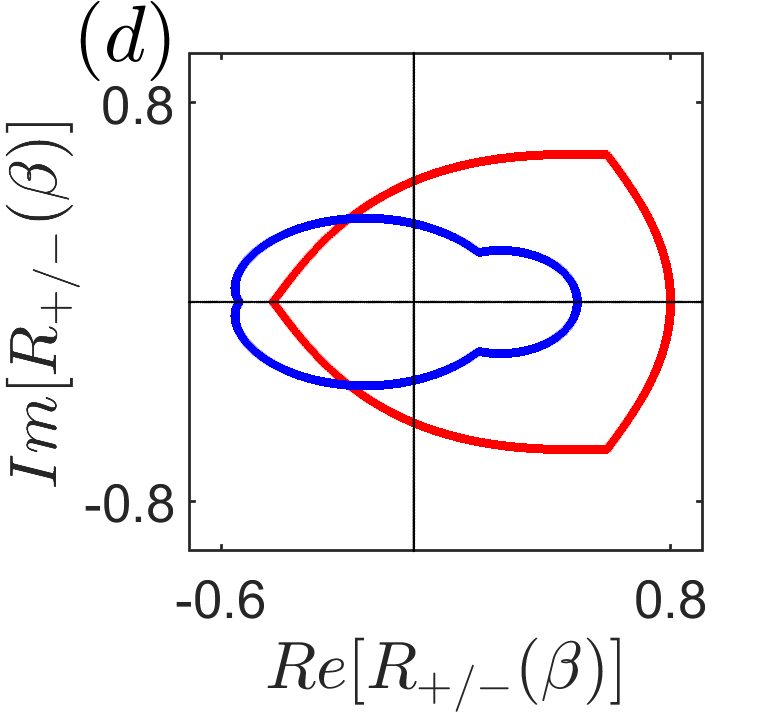}
\includegraphics[width=4.2cm,height=3.8cm]{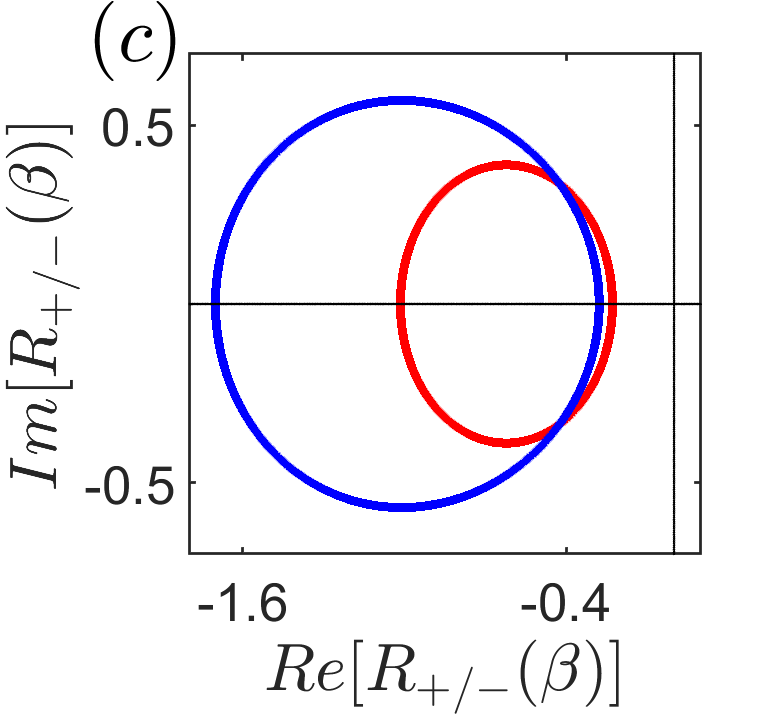}
\caption{(Color online) $N=100$, $\gamma_{1}=0.3$, $t_{2}=\frac{\gamma_{2}}{2}=0.1$, $t_{3}=0.5$, $\gamma_{3}=0.05$. (a) The open boundary energy spectrum. (b)The continuum bands. (c) and (d) show the loops of $R_{+}(\beta)$ (the red line) and $R_{-}(\beta)$ (the blue line) on the complex plane at $t_{1}=0.1$ and $t_{1}=-0.8$, respectively. For (c), both $R_{+}(\beta)$ and $R_{-}(\beta)$ enclose the origin, but for (d), the origin are excluded by $R_{+}(\beta)$ and $R_{-}(\beta)$.}
\label{fig3}
\end{figure}

\begin{figure}[!htbp]
\includegraphics[width=4.2cm,height=3.8cm]{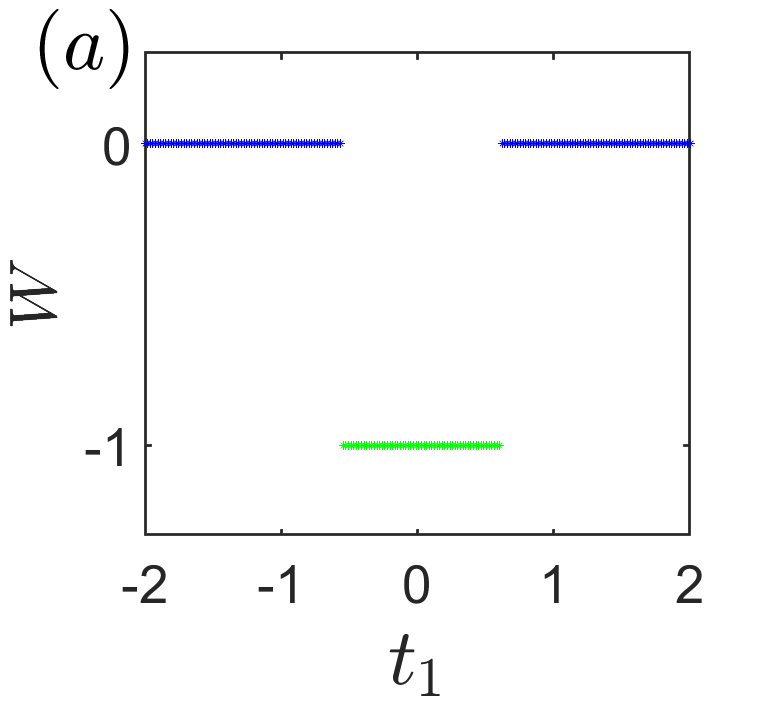}
\includegraphics[width=4.2cm,height=3.8cm]{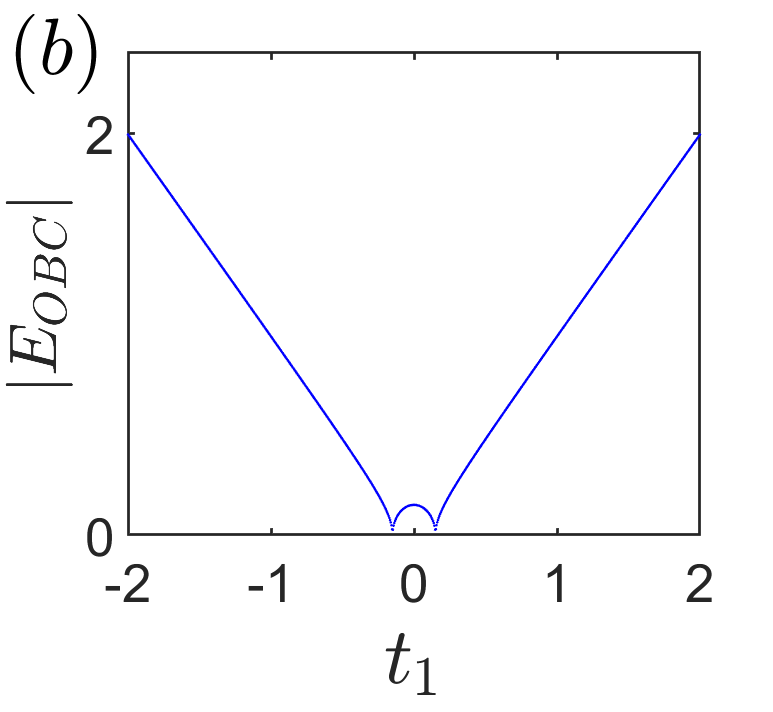}

\includegraphics[width=4.2cm,height=3.8cm]{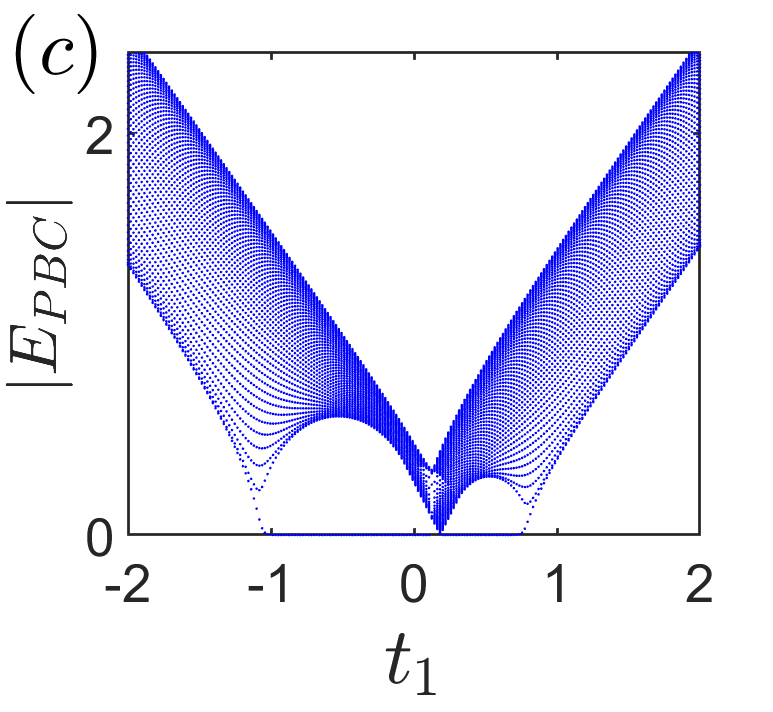}
\includegraphics[width=4.2cm,height=3.8cm]{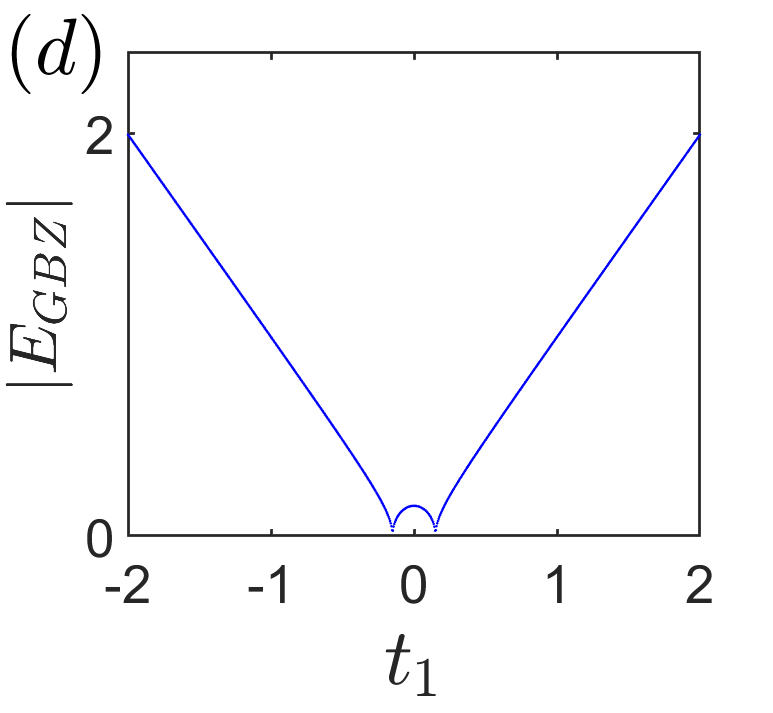}

\caption{(Color online) (a) The topological invariant obtained from the Eq. (8). The parameters are same as Fig. 3(a). (b), (c) and (d) show the breakdown and the recovery of the bulk boundary correspondence from the energy perspective, and the parameters are given by $N=100$, $\gamma_{1}=0.3$, $t_{2}=\frac{\gamma_{2}}{2}=0.1$ and $t_{3}=\frac{\gamma_{3}}{2}=\frac{1}{2}$.}
\label{fig4}
\end{figure}

It can be found that Eq. \eqref{14} has a constant solution of $\beta$=0 and thus the rest of the solutions should be taken precedence. As shown in Figs. \ref{fig3}(a) and \ref{fig3}(b), the open boundary energy spectrum matches well with the continuum bands if the zero energy modes is excluded, i.e., the bulk boundary correspondence can be well revised from the energy side, even one of the solutions is zero for a quartic equation. Furthermore, in Figs. \ref{fig3}(c) and \ref{fig3}(d), both $R_{+}(\beta)$ and $R_{-}(\beta)$ form the closed loop containing the origin or not, which implies $W$ =$1$ and $W$=$0$ for $t_{1}$=$0.1$ and $t_{1}$=$-0.8$, respectively. This result also can be confirmed by the zero modes in the open boundary energy spectrum in Fig. \ref{fig3}(a). Namely, the winding number can be well defined and the system can possess a nontrivial phase in a range of parameters, as shown in Fig. \ref{fig4}(a). Then, the bulk boundary correspondence also can be well reopened from the topological invariant under one of the $\beta$ is zero.

Except for $t_{2}=\frac{\gamma_{2}}{2}$, we assume $t_{3}=\frac{\gamma_{3}}{2}$ additionally. The characteristic equation \eqref{14} can be reduced to
\begin{equation}
\beta^{2}[t_{1}^{2}-\frac{\gamma_{1}^{2}}{4}+2\Big(t_{2}(t_{1}+\frac{\gamma_{1}}{2})+t_{3}(t_{1}-\frac{\gamma_{1}}{2})\Big)\beta+4t_{2}t_{3}\beta^{2}-E^{2}]=0,
\end{equation}
and one can obtain
\begin{equation}
R_{+}(\beta)=t_{1}+\frac{{\gamma_{1}}}{2}+2t_{3}\beta,
\end{equation}
\begin{equation}
R_{-}(\beta)=t_{1}-\frac{{\gamma_{1}}}{2}+2t_{2}\beta.
\end{equation}

Significantly, this quartic equation has multiple roots of $\beta_{1}=\beta_{2}=0$. Inevitably, the GBZ will become a point because of the necessary condition $\left|\beta_{2}\right|=\left|\beta_{3}\right|=0$. Accordingly, the continuum bands can be represented as $E_{GBZ}^{2}=t_{1}^{2}-\frac{\gamma_{1}^{2}}{4}$. In this specified case, the open boundary energy spectrum even can be reconstructed by the continuum bands loyally, but not the periodical boundary energy spectrum, as shown in Figs. \ref{fig4}(b)-\ref{fig4}(d). However, owning to the fact that $R_{+}(\beta)$ and $R_{-}(\beta)$ become the constant in this case, the well definition of the topological invariant can not be ensured. Therefore, if GBZ possesses the bizarre natures, the correctness of the bulk boundary correspondence depends on the point where we interrogate it from.

\section{The illness of the Bulk Boundary Correspondence with The Generalized Brillouin Zone being a closed loop}\label{IV}

\begin{figure}[!htbp]
\includegraphics[width=2.7cm,height=3.8cm]{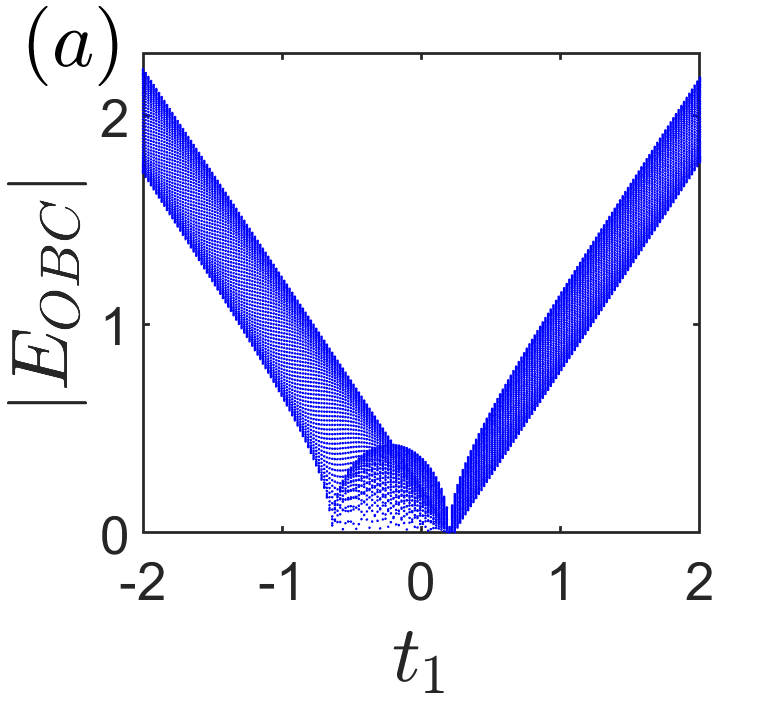}
\includegraphics[width=2.7cm,height=3.8cm]{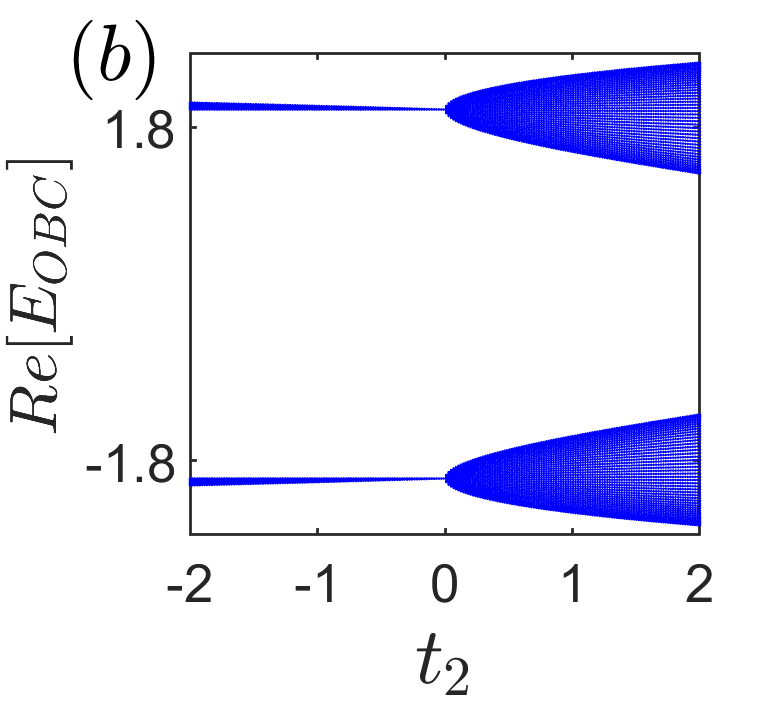}
\includegraphics[width=2.7cm,height=3.8cm]{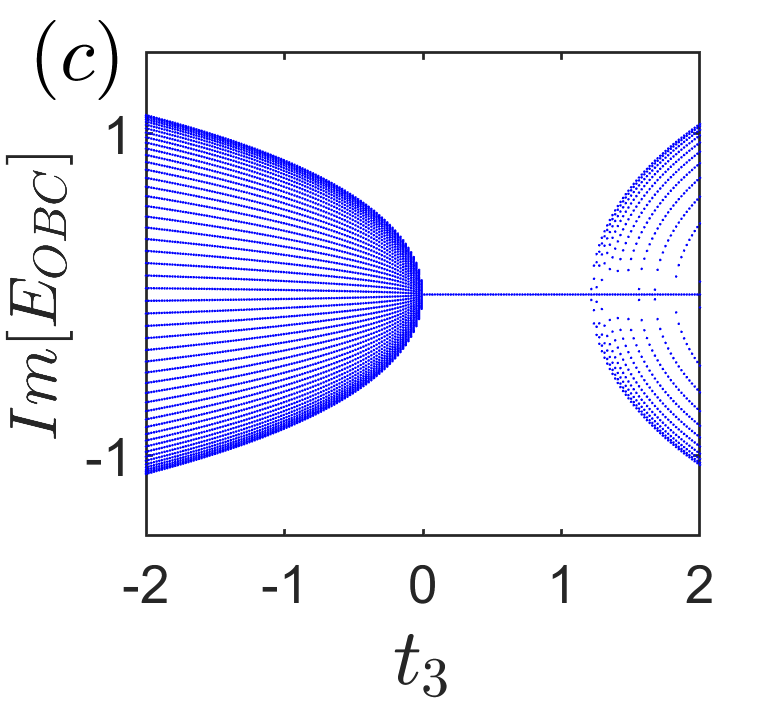}

\includegraphics[width=2.7cm,height=3.8cm]{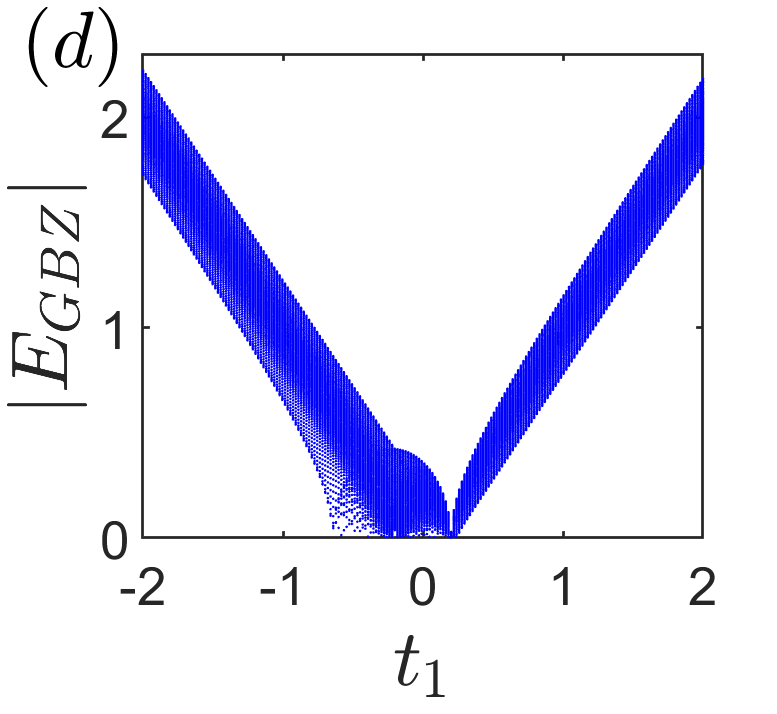}
\includegraphics[width=2.7cm,height=3.8cm]{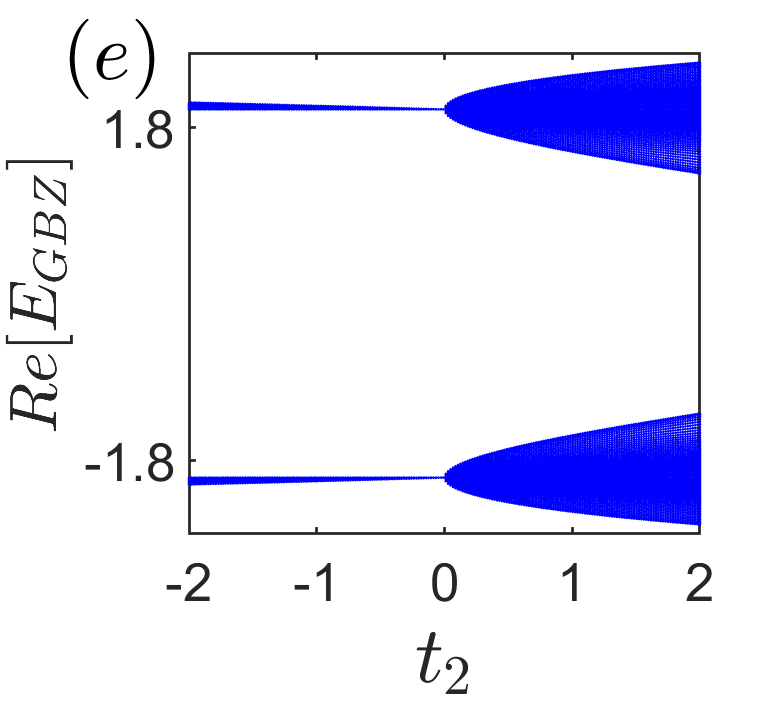}
\includegraphics[width=2.7cm,height=3.8cm]{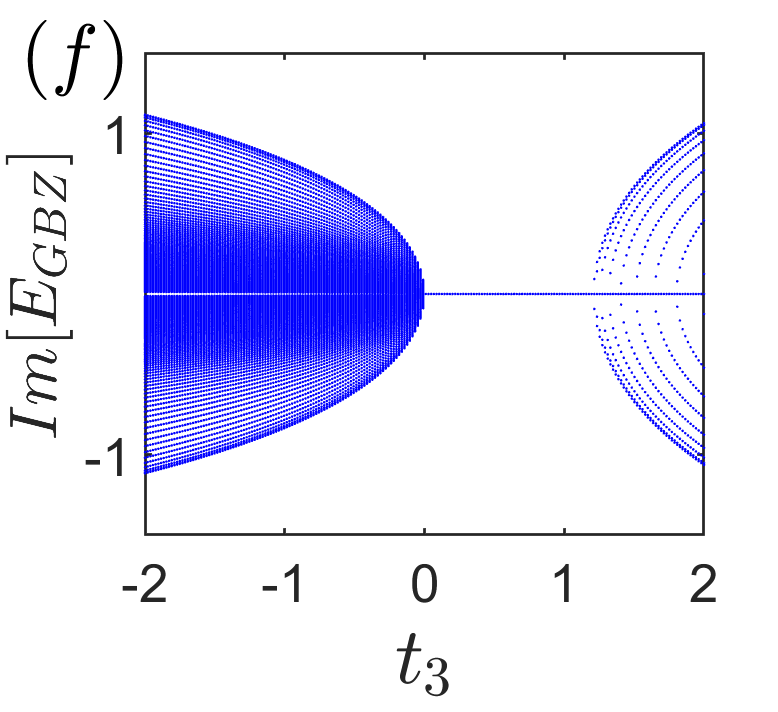}

\caption{(Color online) $N=100$. For (a) and (d), $\gamma_{1}=0.4$, $t_{2}=-\frac{\gamma_{2}}{2}=0.25$ and $t_{3}=\frac{\gamma_{3}}{2}=0.05$. For (b) and(e), $t_{1}=2$, $\gamma_{1}=0.4$ and $t_{3}=\frac{\gamma_{3}}{2}=0.05$. For (c) and (f), $t_{1}=2$, $\gamma_{1}=0.4$ and $t_{2}=-\frac{\gamma_{2}}{2}=0.25$. From (a) to (c), these figures show the open boundary spectra under different cases, which is consistent well with the continuum bands [(d)-(f)] based on the Eq. (19).}
\label{fig5}
\end{figure}

Intuitively, it seems that the illness of the topological number, which is the bone of the bulk boundary correspondence, is induced by the bizarreness of the GBZ. However, we find that even when the GBZ arise with a closed loop exactly, the topological invariant also may be illness. We here take $t_{2}=-\frac{\gamma_{2}}{2}$, which respects
\begin{equation}
[2t_{2}+(t_{1}+\frac{\gamma_{1}}{2})\beta+(t_{3}+\frac{\gamma_{3}}{2})\beta^{2}] [(t_{3}-\frac{\gamma_{3}}{2})+(t_{1}-\frac{\gamma_{1}}{2})\beta]=E^{2}\beta^{2}.
\end{equation}
This is a cubic equation of $\beta$. Then, the bulk boundary correspondence is natural to be renewed as long as the condition $\left|\beta_{2}\right|=\left|\beta_{3}\right|$ is met. However, if one of the three solutions is $\beta=0$, by, e.g., $t_{3}=\frac{{\gamma_{3}}}{2}$, the situation will become subtle,

\begin{equation}
E^{2}_{GBZ}=(t_{1}-\frac{{\gamma_{1}}}{2})(t_{1}+\frac{{\gamma_{1}}}{2})\mp\sqrt{\frac{16t_{2}t_{3}(t_{1}-\frac{{\gamma_{1}}}{2})^{2}}{1+\eta^{2}}},\label{19}
\end{equation}

\begin{equation}
\left|\beta\right|=\sqrt{\left|\frac{t_{2}}{t_{3}}\right|},\label{20}
\end{equation}

\begin{equation}
R_{+}(\beta)=(t_{1}+\frac{{\gamma_{1}}}{2})+2t_{2}\beta^{-1}+2t_{3}\beta,\label{21}
\end{equation}
and
\begin{equation}
R_{-}(\beta)=t_{1}-\frac{{\gamma_{1}}}{2}.\label{22}
\end{equation}
In Fig. \ref{fig5}, three different circumstances of open boundary spectra and the continuum energy depending on Eq. \eqref{19} have been revealed. Evidently, the spectra of the open chain still can be regained faithfully using the $E_{GBZ}$. Essentially, the rehabilitation is reasonable owing to the fact that the Eq. \eqref{20} indicates that the GBZ itself is a circle with the radius of $\left|\beta\right|$, which is a well-defined closed loop in this situation. Note that the elements of the topological invariant of $R_{+}(\beta)$ is the function of GBZ [Eq. \eqref{21}], which also appears in the shape of a closed curve. But, unexpectedly, Eq. \eqref{22} interprets that the another elements of $R_{-}(\beta)$ is just a point in the complex plane even through the GBZ contains finite area. In other words, even if the GBZ is a exactly closed loop, the correctness of the bulk boundary correspondence is only partially established.

\begin{figure}[!htbp]
\includegraphics[width=4.2cm,height=3.8cm]{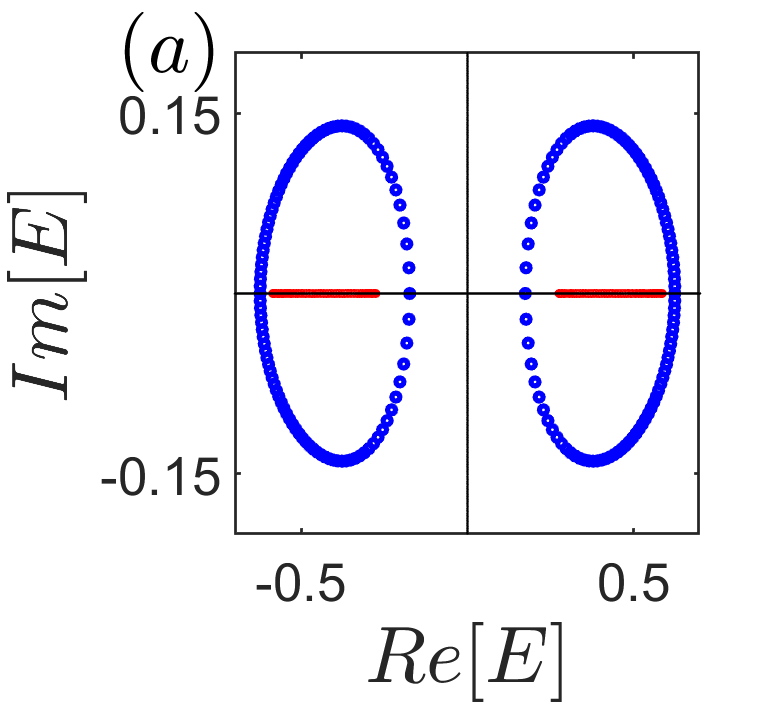}
\includegraphics[width=4.2cm,height=3.8cm]{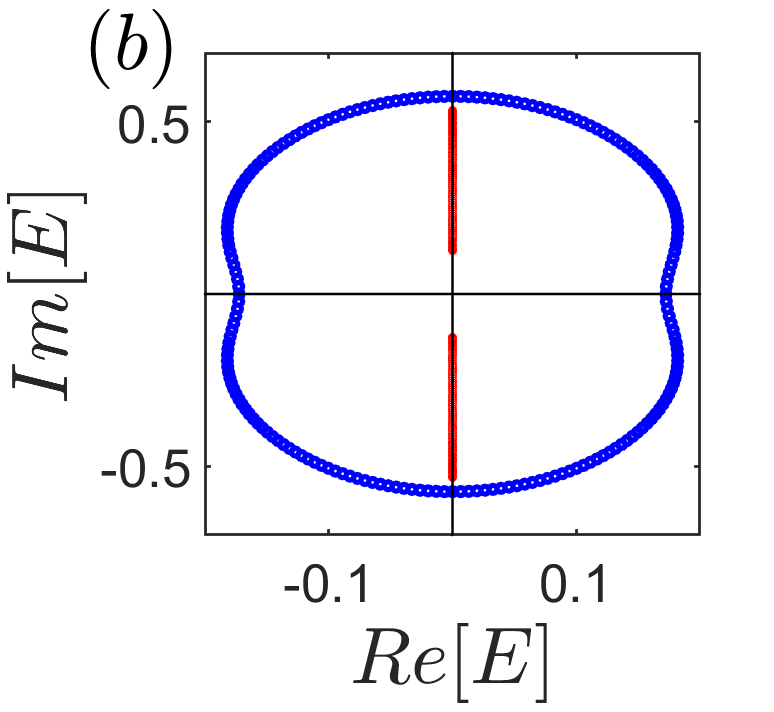}

\caption{(Color online) The common parameters are given by $N=100$, $t_{2}=-\frac{\gamma_{2}}{2}=0.25$ and $t_{3}=\frac{\gamma_{3}}{2}=0.05$. For (a) and (b), $t_{1}=0.5$, $\gamma_{1}=0.4$ and $t_{1}=-0.1$, $\gamma_{1}=-0.8$, respectively. The red line stands for the open boundary spectrum, which is wholly real and imaginary and included by the periodic boundary spectrum (the green line).}
\label{fig6}
\end{figure}

Funnily, the continuum bands of Eq. \eqref{19} induced by $\beta=0$ also tells us an interesting physical properties of the system, i.e., there exists a clear border to distinguish $min[E^{2}_{GBZ}]>0$ or $max[E^{2}_{GBZ}]<0$, by which the eigenvalues of the system is real or purely imaginary can be ensured under the thermodynamic limit for this non-Hermitian system. For example, when we set $t_{1}=0.5$, $\gamma_{1}=0.4$ and other parameters are same as Fig. \ref{fig5}(a), the result of $min[E^{2}_{GBZ}]>0$ implies the open boundary spectrum is completely real.  The numerical result in Fig. \ref{fig6}(a) tests our prediction. Similarly, the parameters $t_{1}=-0.1$ and $\gamma_{1}=-0.8$ will lead to the $max[E^{2}_{GBZ}]<0$, i.e., the open boundary spectrum is totally imaginary, as expected, which also can be identified numerically [Fig. \ref{fig6}(b)].



\section{Conclusion}\label{V}
In this work, a one dimensional non-Hermitian model has been constructed to investigate the physical properties of the bulk boundary correspondence under the circumstance that the GBZ has bizarre features. We have calculated both the continuum bands and topological invariant relevant to the GBZ and the energy spectra under different boundary conditions. It can be found that the energy band of the open chain always can be regained by the continuum bands $E_{GBZ}$, no matter if the GBZ is a point or not. Oppositely, the singularity of the GBZ will cause the ill-definition of the topological invariant. Accordingly, the bulk boundary correspondence ever can be recovered if we only restrict this concept to band structures, but not relate to the topological number. Counter-intuitively, we also find that the bulk boundary correspondence may retain illness even though the GBZ is a closed loop, since the elements of the invariant can be a constant. Moreover, those discoveries are an effective supplement to the current non-Bloch band theory.

\section{ACKNOWLEDGMENTS}
This work was supported by NSFC under grants No.11874190.
\bibliography{ref}

\end{document}